\documentclass[review]{elsarticle}

\usepackage{lineno,hyperref}
\modulolinenumbers[5]

\journal{Journal of \LaTeX\ Templates}

\begin{document}

\begin{frontmatter}

\title{Different Non-extensive Models for heavy-ion collisions}

\author[author1]{K.~M.~Shen}
\ead{shenkm@mails.ccnu.edu.cn}

\author[author2]{T.~S.~Bir\'o}
\ead{Biro.Tamas@wigner.mta.hu}

\author[author1]{E.~K.~Wang}

\address[author1]{Key Laboratory of Quark $\&$ Lepton Physics and Institute of Particle Physics, Central China Normal University, Wuhan}
\address[author2]{Institute for Particle and Nuclear Physics, Wigner Reserach Centre for Physics, Hungarian Academy of Sciences, Budapest}

\begin{abstract}
The transverse momentum ($p_T$) spectra from heavy-ion collisions at intermediate momenta are
described by non-extensive statistical models.  Assuming a fixed relative variance of the
temperature fluctuating event by event or alternatively a fixed mean multiplicity in a negative
binomial distribution (NBD), two different linear relations emerge between the temperature, $T$, and
the Tsallis parameter $q-1$. Our results qualitatively agree with that of G.~Wilk.
Furthermore we revisit the ''Soft+Hard'' model, proposed recently by G.~G.~Barnaf\"oldi
\textit{et.al.}, by a $T$-independent average $p_T^2$ assumption. Finally we compare results
with those predicted by another deformed distribution, using Kaniadakis' $\kappa$ parametrization.
\end{abstract}

\begin{keyword}
non-extensive; heavy-ion collisions; Tsallis parameter $q$; negative binomial distributions; Kaniadakis' kappa-distribution
\end{keyword}

\end{frontmatter}

\section{Introduction}

Non-extensive statistics, as the simplest and most natural generalization of the canonical
Boltzmann--Gibbs statistics, can be applied widely. Heavy-ion collisions also offer an excellent
field for such applications.
The transverse momentum spectra exhibit power-law-tailed behavior, 
that can be described with non-extensive distribution functions 
better than with  Boltzmann distributions, $\exp(-\beta E)\to \exp_q(-\beta E)$. 
Here the notion of $q-$exponential, $e_q(x)$, stands for 
  \begin{eqnarray}
  e_q(x) :=\: [1+(1-q)x ]^{\frac{1}{1-q}}.
  \label{q-ex}
  \end{eqnarray}
This form was first suggested by V.~Pareto in 1896 \cite{Pareto-1896} for describing 
the distribution of wealth, and it has been recently promoted by C.~Tsallis \cite{Tsallis-1988} 
in connection with non-extensive entropy.
The use of a cut power-law in high energy physics was first introduced by R.~Hagedorn \cite{Hagedorn-1984},
describing the data of the invariant cross section of hadrons as a function of $p_T$ over a wide range.

In order to gain an improved insight into this generalized non-extensive statistics, as well as
to value its application to $p_T$ spectra on the top of pQCD motivated power-laws,
ever used for high-energy jets, the meaning of its key parameter, $q$, must inevitably be studied. 
Considering the thermodynamical properties of hot and dense nuclear matter formed in heavy-ion collisions, 
on the other hand it is worth to study the fluctuations of the relevant physical observables,
with a large number of particles produced. For example the multiplicity fluctuations 
in heavy-ion collisions have been investigated by several authors, including 
G.~Wilk \textit{et al.}\cite{fluctuation-2009}.
Connecting to such studies, we have obtained Tsallis' distribution for particular particle 
number fluctuation patterns due to a finite heat bath, 
for details see \cite{Biro-2015}. 
 
On the other hand, further non-extensive distribution functions were being proposed and studied
to describe the data on $p_T$ spectra in heavy-ion collisions. 
K.~Urmossy \textit{et al.} proposed a model in which the hadrons produced in heavy-ion collisions 
stem from a quark-gluon plasma (QGP), referred to as 'soft' yields, on top of those
stemming from jets, called 'hard' yields \cite{Urmossy-2014}. 
Both of these contributions are distributed according to respective cut power-laws,
$\exp_q(-\beta E)$. Based on this view in this paper we investigate the scatter of the parameter 
$q$ with respect to the fitted temperatures $T$. This investigation should differentiate
between hard and soft origin of power laws and should look different for heavy-ion as for
$pp$ data.
 
Considering quantum-statistics even further constraints arise from particle-hole CPT symmetry.
They lead to a special requirement on the cut power-law function, alternatively called deformed
exponential \cite{KMS-2015}:
 \begin{equation}
 e_q(x)\cdot e_q(-x)=1.
 \label{KMS-e2015}
 \end{equation}
For the Pareto-Hagedorn-Tsallis distribution, $\exp_q(x)$,
it is easy to see that it does not satisfy this relation, therefore it is
inadequate to apply naively extended quantum statistical distributions in the form
\begin{eqnarray}
 n_{B,F}=\frac{1}{e_q(x)\mp 1}
 \label{number-e1}
\end{eqnarray} 
with $x=(\omega -\mu)/T$, where $\omega$, $\mu$ and $T$ are the energy, chemical potential
and temperature of the system, respectively, following the notation used in \cite{number-3}.
In the present paper we also shall apply the $\kappa$-exponential distribution, 
proposed by G.~Kaniadakis for the description of relativistic plasmas \cite{kappa-2001}. 
We compare our corresponding analysis with the ones using Tsallis' distribution, discussed above. 
The $\kappa$-exponential function has the advantage that it 
readily satisfies the particle-hole symmetry demand:
\begin{eqnarray}
 e_{\kappa}(x)\cdot e_{\kappa} (-x)=1,
 \label{KMS-2015-1}
\end{eqnarray}
based on the definition
\begin{eqnarray}
 e_{\kappa}(x) : = \:  [\sqrt{1+(\kappa x)^2}+\kappa x]^{\frac{1}{\kappa}}.
 \label{kappa-2001-e1}
\end{eqnarray}


\section{$p_T$ spectra with finite heat capacity}

Particle transverse momentum ($p_T$) spectra provide a tool for measuring thermal
properties of the QGP formed in ultra-relativistic heavy-ion collisions.
Nowadays more and more experimental data show that the $p_T$ spectrum
exhibits power-like rather than the earlier expected exponential behavior, and multiparticle
distributions are also broader than expected \cite{Wilk-2012}. The number of particles in such
processes is not that large ($N \sim 10^3-10^5~\ll 10^{23}$, much less than 
Avogadro's number). 
All of these warrant a suitable modification of the thermal model by accounting
for possible intrinsic fluctuations in observables measured in high-energy collisions.
 
The non-extensive form of statistical mechanics, proposed by Tsallis \cite{Tsallis-1988},
has found applications in many fields including high energy physics \cite{ALQ-2000,DBSB-2007}.
In several situations the $p_T$ spectra can be best described by a Tsallis distribution, characterized by
a non-extensive parameter $q$ and a scale parameter $T$ \cite{Wong-2013}. 
G. Wilk \textit{et al.} have demonstrated that $q$ turns to be a function of fluctuations of temperature $T$,
together with fluctuations of other variables, from an analysis of different experimental 
data \cite{Wilk-2015}.
Moreover, considering the statistical power-law tailed distributions as canonical distributions
in a thermal system connected to a heat reservoir with finite heat capacity\cite{Biro-2015}, we have
the combined expression 
\begin{eqnarray}
  q=1-\frac{1}{C}+\frac{\Delta T^2}{T^2},
  \label{q-heat}
  \end{eqnarray} 
as next to leading order approximation in the total size of the system.
Here $C$ denotes the heat capacity of the reservoir, and $1/C=dT/dE$.
It is connected with the temperature and the underlying equation of state, $S(E)$,
of the system via the relations $\langle S^{\prime}(E) \rangle = 1/T$ and
$\langle S^{\prime\prime}(E) \rangle = -1/CT^2$. Here the averaging is done
over the ensemble of different collision events with fluctuating parameters.
For temperature fluctuations with the Gaussian variance, $\Delta T/T=1/\sqrt{C}$, 
following from traditional consideration, one arrives at $q=1$: nothing but the classical
Boltzmann--Gibbs case.
  
\begin{figure}[!htb]
\centerline{
	\includegraphics[width=100mm,angle=0,scale=1]{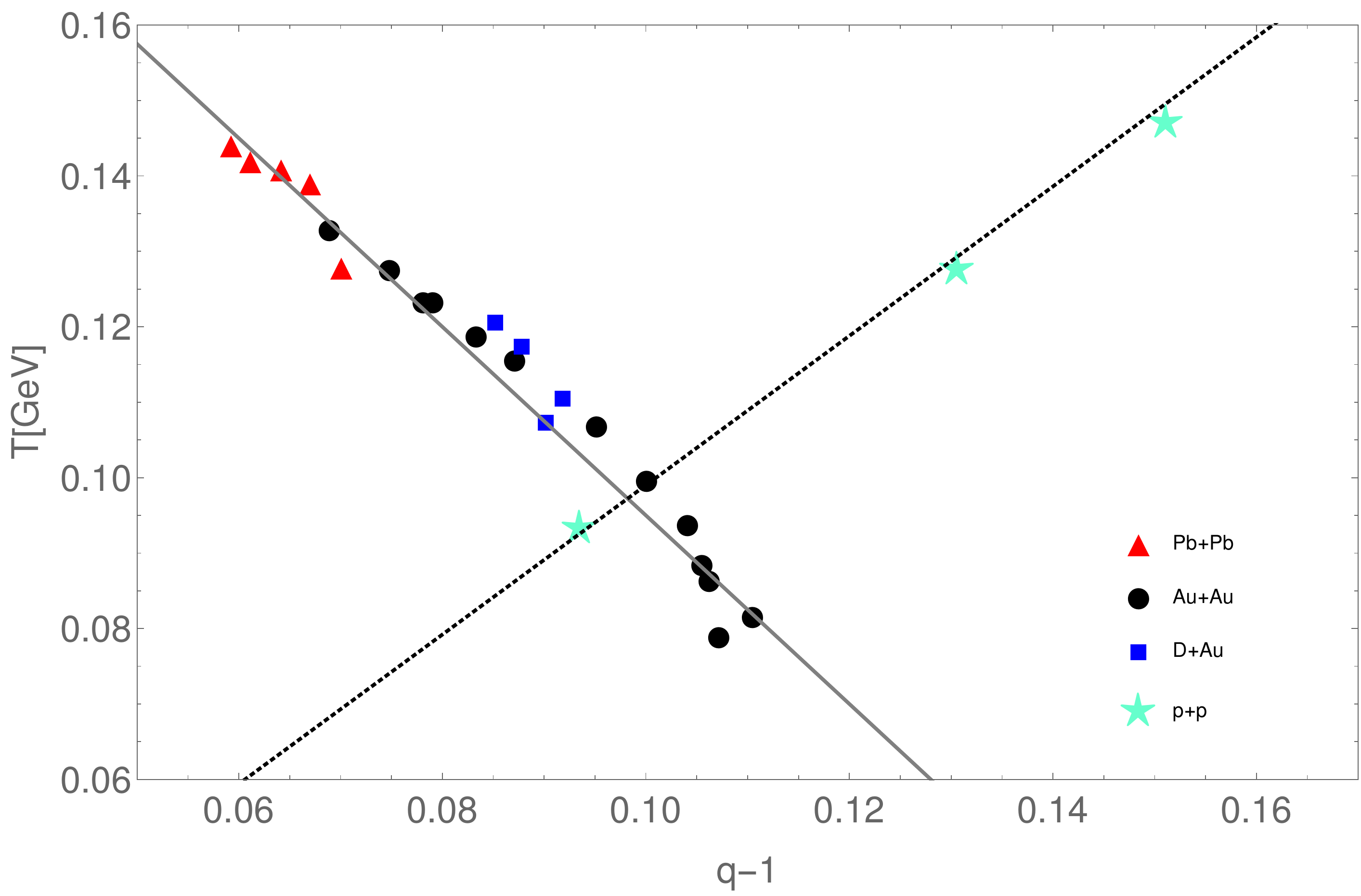}%
	}
\caption{(Color Online) Data are from G. Wilk's collection in his talk given 
at the Erice School on Complexity, 2015\cite{Wilk-2015}. 
our fitting plot is shown with the black line, $T=0.22-1.25(q-1)$ GeV for the constant relative variance 
	$\sigma ^2$ case ($E=1.25$ and $\sigma ^2=0.176$) following Eq.(\ref{Const-Var}) and
	the dotted one 
	$T=q-1$ GeV for the constant average occupancy $f$ case ($E/f=0.99$) from Eq.(\ref{Const-Occ}), respectively.}
\label{constCplot}
\end{figure}

Based on this more generalized function of the Tsallis parameter $q$, we consider two different situations:  
\begin{enumerate}  
\item Constant Relative Variance due to $\beta=S^{\prime}(E)$-Fluctuations.
In multi-particle production processes in heavy-ion collisions, it has 
been shown that the local temperature, $1/\beta$,
fluctuates from point to point around some equilibrium value, $T = \langle 1/\beta \rangle$.
However, for different collisions the relative variance due to its fluctuations stays
approximately constant, namely
\begin{equation}
	\sigma^2 =\frac{\Delta T^2}{T^2}\approx {\rm const}.
   \label{Const-Variance}
   \end{equation}
Thus we conclude from Eq.(\ref{q-heat}) that
\begin{equation}
   T=E[\sigma^2 -(q-1)]
   \label{Const-Var}
\end{equation}
where we approximate the finite heat capacity by the ideal gas ratio $C=E/T$.
\item Constant Average Occupancy.
For elementary $pp$ collisions on the other hand, the measurements of charged particle 
multiplicity distributions are well
described by the negative binomial distribution (NBD) for the full rapidity region.\cite{DM-2010}
Within the NBD scenario we have the probability
\begin{equation}
   P_n=\left( \begin{array}{c} n+k \\ n \\ \end{array} \right) f^n(1+f)^{-n-k-1}
\end{equation}
therefore we obtain the generating function
\begin{equation}
  G(t) \: = \: \sum P_n e^{nt} \: = \: -(k+1)\ln \left(1+f - fe^{t} \right).
\label{Gt-NBD}
\end{equation}
For the expectation value and variance in the number we get
\begin{eqnarray}
 \langle n \rangle &=& G'(0) \: = \: (k+1)\, f
 \nonumber \\
 \Delta n^2 &=&  G''(0) \: = \: (k+1)\, f(f+1).
\label{Gt-NBD-2}
\end{eqnarray}
Assuming now that the average occupancy of phase space by the newly produced
hadrons, $f$,  is constant, 
we arrive at another relation between $T$ and $q$:
   \begin{equation}
   T=\frac{E}{f}(q-1).
   \label{Const-Occ}
   \end{equation}
Again we used relatrions valid for ideal gases,
$C=E/T = \langle n\rangle$ and $\Delta T^2/T^2=\Delta n^2/\langle n\rangle ^2$.
\end{enumerate}   
In Fig.\ref{constCplot} the comparison between our model predictions 
and G Wilk's collection is shown. From it we conclude that
(i) For the fitting parameters with a Tsallis distribution in $AA$ collisions, assuming constant relative
variance of temperature for different collisions works well with experimental data. 
(ii) For $pp$ collisions the relative fluctuations are no longer the same but the average occupancy 
parameter in the NBD looks constant for the full
rapidity region. This makes another line of correlation between $T$ and $q$ as seen in 
Fig.\ref{constCplot}.  There is a remarkable agreement between our simple model predictions and G. Wilk's
extensive data fits.

\section{$p_T$ spectra within A 'Soft+Hard' Model}
 
Recently a so-called 'soft+hard' model was proposed in Ref.\cite{Urmossy-2014}.
According to this model the $p_T$ spectra are divided into
hadron yields stemming from a supposedly 3d-isotropic QGP (soft) and from quasi
1d-jets (hard), respectively:
\begin{equation}
 p^0\frac{dN}{d^3p}=(p^0\frac{dN}{d^3p})^{hard}+(p^0\frac{dN}{d^3p})^{soft}.
 \label{e1}	
\end{equation}
For both yields the classical Tsallis distribution, $e_q(-\omega /T)$, has been applied
with independently fitted $T$ and $q$ parameters \cite{model-2015}.
 
We apply in the present work a radial blast wave flow based
Doppler factor to the heavy-ion data, interpreting the spectral temperature as
\begin{equation}
 T : =\: T_D \sqrt{\frac{1-v}{1+v}},
 \label{Doppler-T}
\end{equation}
with $T_D$ Doppler-shifted Temperature.
This is a simplification compared to the classical Tsallis distribution,
that interprets the single particle energy as a relativistic expression with a given mass, $m$:
\begin{equation}
 \omega : =  \gamma (m_T - vp_T) - m,
\end{equation}
with $m_T=\sqrt{m^2+p_T^2}$ transverse mass and $\gamma=1/\sqrt{1-v^2}$. For $p_T \gg m$
these expressions agree.
Then we inspect the $p_T$ spectra of charged hadrons as a sum of hard and soft yields
 \begin{equation}
 p^0\frac{dN}{d^3p} = \sum_i C_i \cdot e_{q_i}(-\omega_i/T_i)
 \label{pt-1}
 \end{equation}
for $i=soft, ~hard$ and the $C_i$-s being normalization constants.
 
\begin{figure}[!htb]
\centerline{
 \includegraphics[width=100mm,angle=0,scale=1]{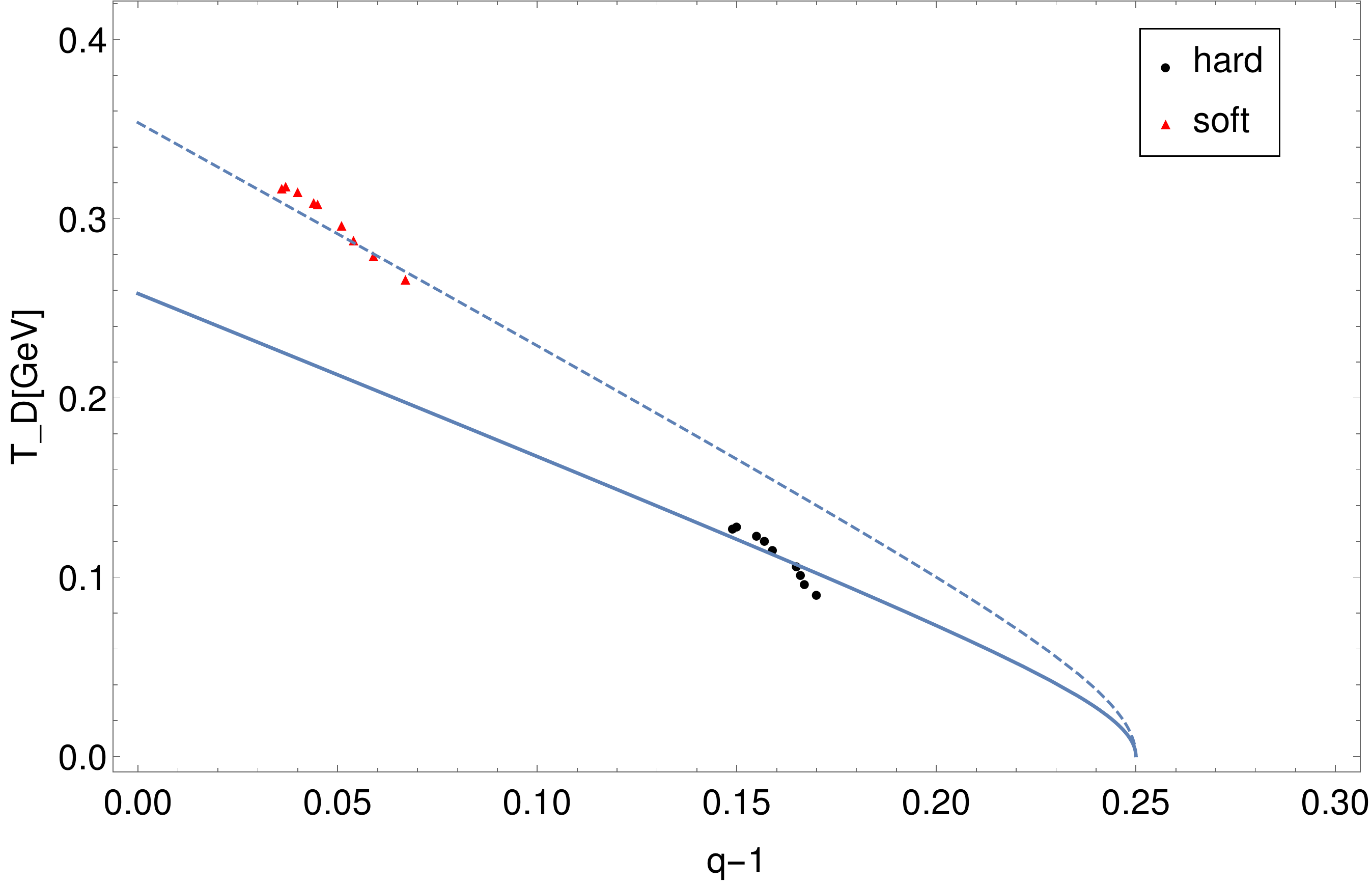}%
	}
\caption{Parameters are from fitting the Transverse momentum spectra of charged pions stemming 
from different centralities of $PbPb$ collisions at $\sqrt{s}=2.76$ TeV, as well as our
fittings with the 'soft + hard' model using Tsallis-$q$ deformed distributions Eq.(\ref{fit-formula}),
which are compared with our theoretical results, 
and curves are fits of constant $\langle p_T^2\rangle$ values via Eq.(\ref{pt2-q}).}
\label{constpt2plotq}
\end{figure}

For light particles like the pion in the observed $p_T$ range the assumption
$p_T\gg m$ can safely be made and therefore $\omega \rightarrow \gamma (1-v)p_T$ leads to 
\begin{equation}
 \frac{\omega}{T} \: \rightarrow \: \frac{\gamma (1-v)p_T}{T} \: =\: \frac{p_T}{T_D}.
\label{Doppler-x}
\end{equation}
Investigating pion yields from various centrality classes we fit therefore
a sum of soft and hard $q-$deformed distributions using the variable $x=p_T/T_D$:
\begin{equation}
	p^0\frac{dN}{d^3p} \Bigg|_i \: = \: C_i \cdot e_{q_i}(-p_T/T_{Di}).
\label{fit-formula}
\end{equation}
Moreover, for $p_T$ spectra stemming from the same identified hadrons but just for different centralities,
the average $p_T^2$ can also be considered as a constant; it is connected with nothing but the
collision energy.  In this model approximation we obtain
\begin{equation}
\langle p_T^2\rangle \: \approx \:
\frac{\int p^3_Tdp_T(1+\frac{q-1}{T_D}p_T)^{-\frac{1}{q-1}}}%
	{\int p_Tdp_T(1+\frac{q-1}{T_D}p_T)^{-\frac{1}{q-1}}} 
\: = \:  \frac{6\, T_D^2}{(4-3q)(5-4q)}.
\label{pt2-q}
\end{equation}
In order to test these assumptions we fitted $p_T$ spectra measured in $PbPb$ collisions 
at $\sqrt{s}=2.76$ TeV by the ALICE experiment using the functional form in Eq.(\ref{pt-1}).
We used identified hadron spectra and then the fitted parameters were compared with our results 
from above, cf.  Fig.\ref{constpt2plotq}.
In Fig.\ref{constpt2plotq}, we present the 'soft + hard' model fits for the
$p_T$ spectra of charged pions in $PbPb$ collisions at $\sqrt{s}=2.76$ TeV, as seen in\cite{model-2015}.
The fitted parameter values are listed in Tab.\ref{tab-q}; here fluctuations and 
normalization factors are not explicitly shown, for simplicity . As
for the fitting parameters, we used our theoretical and phenomenological model with the constant 
average $p_T^2$ to describe the connection between $T_{Di}$ and $q_i$. Moreover, it
is demonstrated that parameters fit the theoretical results quite well.
In this fit for the soft part, we obtain $\langle p_T^2\rangle \approx 0.75$ GeV$^2$, 
while $\langle p_T^2\rangle \approx 0.40$ GeV$^2$ for the hard part. 
This also supports that $p_T$ spectra can be studied in two parts, like
in the 'soft + hard' model \cite{Urmossy-2014}.

\begin{figure}[!htb]
\centerline{
	\includegraphics[width=70mm,angle=0,scale=0.8]{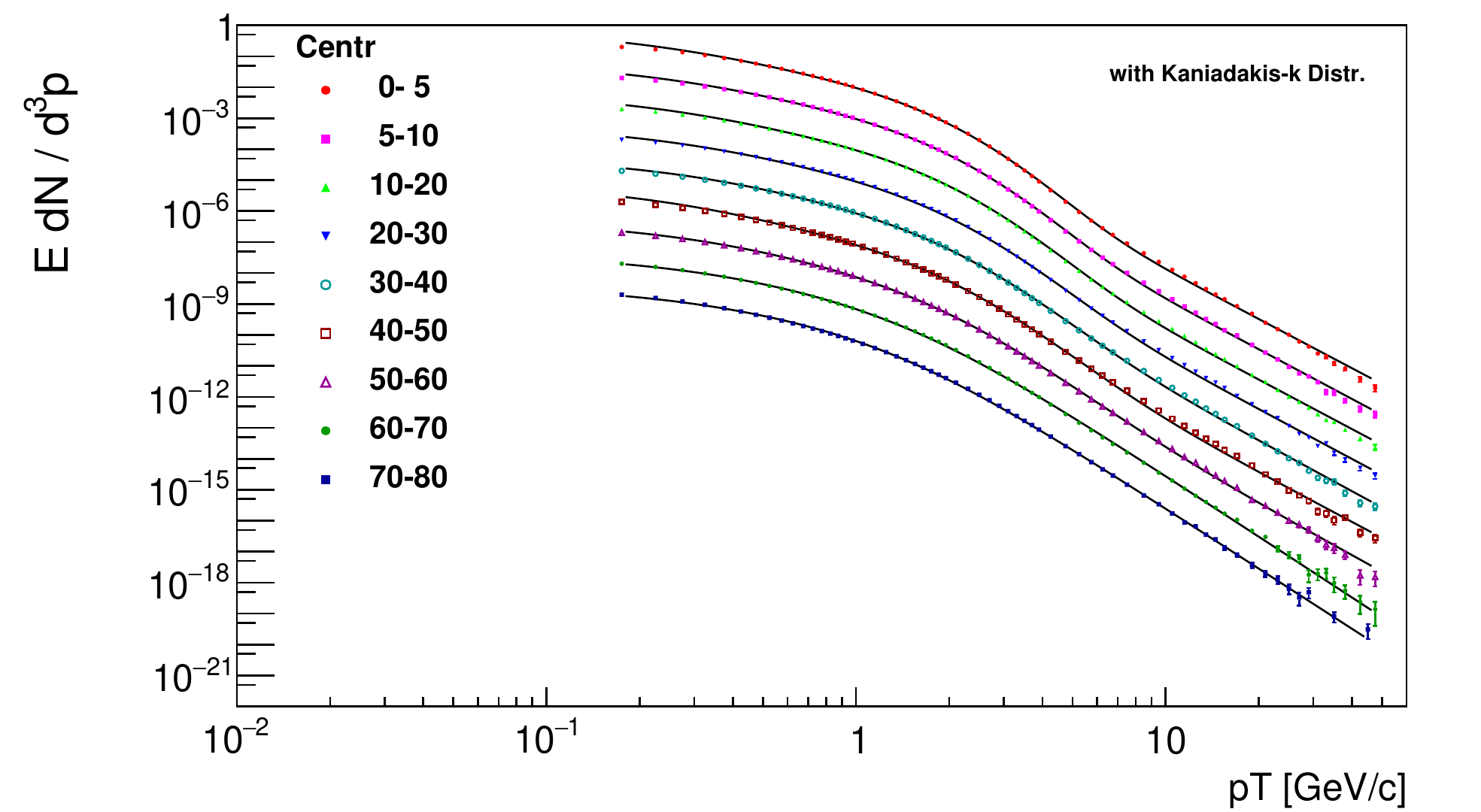}%
	\includegraphics[width=60mm,angle=0,scale=0.8]{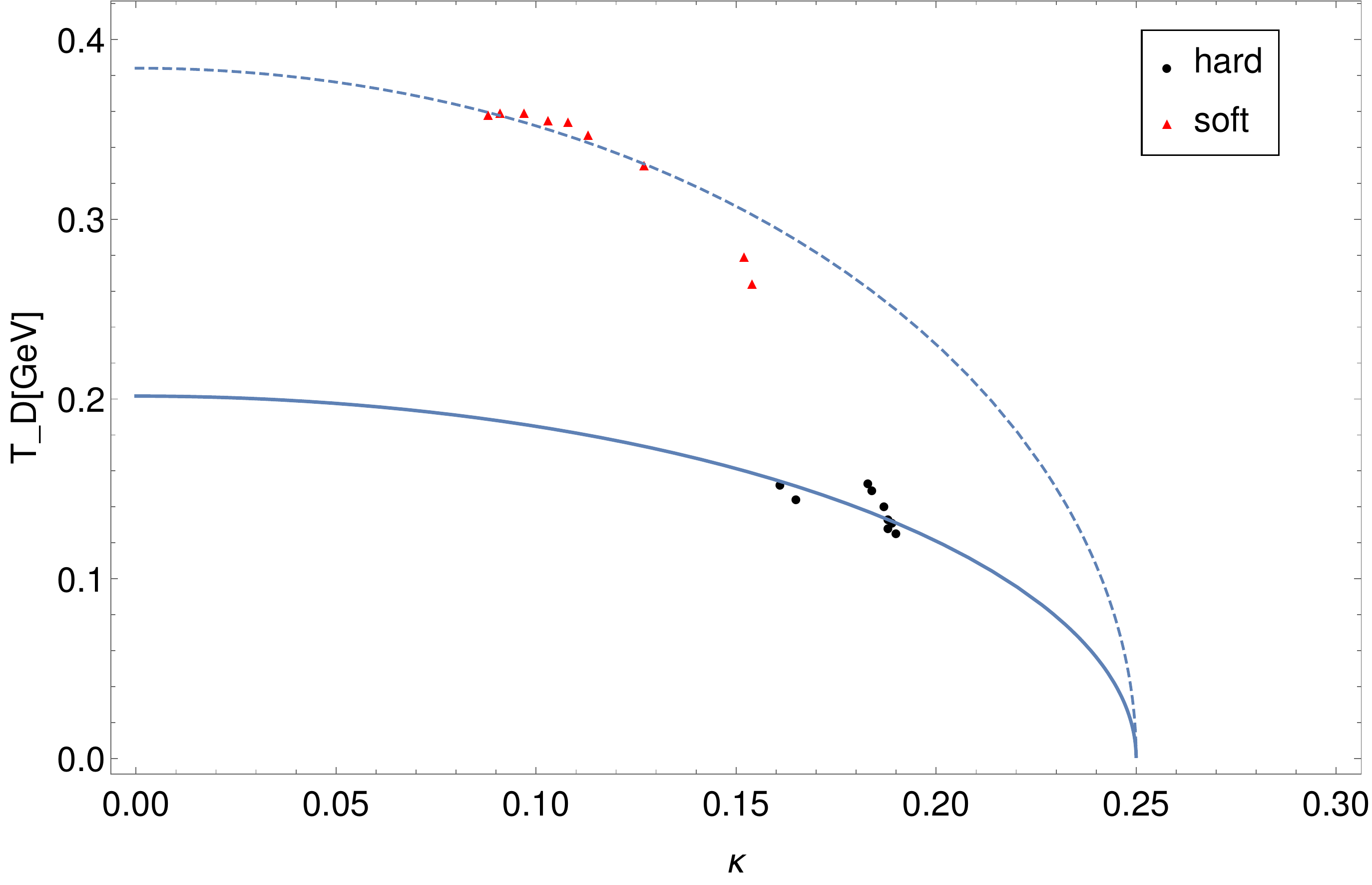}%
	}
\caption{Left panel: Transverse momentum spectra of charged pions stemming 
from different centrality $PbPb$ collisions at $\sqrt{s}=2.76 TeV$, as well as our
fittings with the 'soft + hard' model using $\kappa$- deformed distributions Eq.(\ref{fit-formula-k}). 
Right panel: Fit parameters are compared with our theoretical results: 
the curves are fits of Eq.(\ref{pt2-k}) with the respective $\langle p_T^2\rangle$ values of
$0.88$ GeV$^2$ and $0.24$ GeV$^2$ for the soft and hard part.}
\label{constpt2plotk}
\end{figure}

\begin{table}[!htb]
\caption{Fit parameters to $p_T$ spectra in $PbPb$ collisions within the 'soft + hard' model of Tsallis$-q$.}
\begin{center}
\begin{tabular}{lllll} \hline\noalign{\smallskip}
Centrality, $\:$ Parameters: & $q_{soft}$ & $T_{D,soft}$ & $q_{hard}$ & $T_{D,hard}$ \\
\noalign{\smallskip}\hline\noalign{\smallskip}
$0 \sim 5$ & 1.036 & 0.317 & 1.170 & 0.091 \\
$5 \sim 10$ & 1.037 & 0.318 & 1.167 & 0.097 \\
$10 \sim 20$ & 1.040 & 0.315 & 1.166 & 0.102 \\
$20 \sim 30$ & 1.044 & 0.309 & 1.165 & 0.107 \\
$30 \sim 40$ & 1.045 & 0.308 & 1.159 & 0.116 \\
$40 \sim 50$ & 1.051 & 0.296 & 1.157 & 0.121 \\
$50 \sim 60$ & 1.054 & 0.288 & 1.155 & 0.124 \\
$60 \sim 70$ & 1.059 & 0.279 & 1.150 & 0.129 \\
$70 \sim 80$ & 1.067 & 0.266 & 1.149 & 0.128 \\
\noalign{\smallskip}\hline
\end{tabular}
\end{center}
\label{tab-q}
\end{table}  
In the followings we demonstrate that using Kaniadakis' kappa-distribution, fit for
the non-extensive extension of quantum statistics, works equally well
as the Tsallis one in the 'soft + hard' model, cf. Fig.\ref{constpt2plotk}.
The spectra using the $\kappa$-deformed distribution Eq.(\ref{kappa-2001-e1}) are given as
\begin{equation}
	p^0\frac{dN}{d^3p} \Bigg|_i \: = \: C_i \cdot e_{\kappa_i}(-p_T/T_{Di})
\label{fit-formula-k}
\end{equation}
Then the corresponding $\langle p_T^2\rangle$ is
\begin{equation}
 \langle p_T^2 \rangle \: = \: \frac{6 \, T_D^2}{1-16\kappa^2}.
\label{pt2-k}
\end{equation}
In Fig.\ref{constpt2plotk} it is shown that this model also works quite well with $\kappa$-deformed
distributions. Within the constant $\langle p_T^2\rangle$ the fitting parameters $\kappa_i$ and $T_{Di}$
are related with each other too, their values are listed in Tab.\ref{tab-k}. 
Worth to be mentioned that the total sum of soft and hard contributions to $\langle p_T^2\rangle$ 
should be and is indeed model-independent: $0.75+0.40\approx 0.88+0.24$.
\begin{table}[!hbt]
\caption{Fit parameters of $p_T$ spectra in $PbPb$ collisions with the 'soft + hard' model using
	Kaniadakis' $\kappa$.}
\begin{center}
\begin{tabular}{lllll} \hline\noalign{\smallskip}
Centrality, $\:$ Parameters & $\kappa_{soft}$ & $T_{D,soft}$ & $\kappa_{hard}$ & $T_{D,hard}$ \\
\noalign{\smallskip}\hline\noalign{\smallskip}
$0 \sim 5$ & 0.088 & 0.358 & 0.190 & 0.126 \\
$5 \sim 10$ & 0.091 & 0.359 & 0.188 & 0.134 \\
$10 \sim 20$ & 0.097 & 0.359 & 0.189 & 0.132 \\
$20 \sim 30$ & 0.103 & 0.355 & 0.187 & 0.141 \\
$30 \sim 40$ & 0.108 & 0.354 & 0.184 & 0.150 \\
$40 \sim 50$ & 0.113 & 0.347 & 0.188 & 0.129 \\
$50 \sim 60$ & 0.127 & 0.330 & 0.183 & 0.154 \\
$60 \sim 70$ & 0.152 & 0.279 & 0.161 & 0.153 \\
$70 \sim 80$ & 0.154 & 0.264 & 0.165 & 0.145 \\
\noalign{\smallskip}\hline
\end{tabular}
\end{center}
\label{tab-k}
\end{table}  



 
\section{Conclusions and outlook}
 
In order to interpret the transverse momentum spectrum in heavy-ion collisions in terms
of thermal concepts, we have considered several models based on non-extensive statistical 
mechanics.  Both the original and a  generalized Tsallis-Pareto distributions
were fitted. Firstly, as QGP is a well known candidate for a thermal system, we have reminded that
Tsallis' $q$ parameter is connected to the finite heat capacity of the reservoir as well as 
to the intrinsic fluctuations of temperature.
Then we have demonstrated that for $pp$ and $AA$ collisions the non-extensive parameter $q$ and
temperature $T$ must fulfill different relations. 
Different formulas have been derived and both results are well fitted with 
other phenomenological results, like G. Wilk's data collection shown
in a talk at the Erice School on Complexity, in 2015 \cite{Wilk-2015}.
 
As a further approach we have considered a constant $\langle p_T^2 \rangle$
for various centrality collisions,
and used this model to investigate the collection of fit parameters $q$ and $T$.
Based on experimental data taken at different centralities in $PbPb$ collisions at $\sqrt{s}=2.76$ TeV, 
we fitted the 'soft + hard' model to $p_T$ spectra. The Tsallis' $q$ is found to be well 
connected with the temperature, $T$, gathering along lines following from simple
arguments. 
 
We have also investigated this correspondence using another deformed distribution, 
Kaniadakis' $\kappa$-distribution. We have demonstrated
that the 'soft+hard' model works equally well with this deformed exponential function. 
This is important, because quantum statistical distributions, like the Fermi and Bose
distributions, can be constructed by using this particle--hole symmetric deformation
of the exponential easily. In the future we plan to explore the applicability
of the $\kappa$-distribution to detailed spectra stemming from ultra-relativistic
heavy-ion collisions.


{\em Acknowledgement}
 This work has been supported in part by MOST of China under 2014DFG02050,
 the Hungarian National Research Fund OTKA (K104260), a bilateral governmental
 Chinese-Hungarian agreement NIH TET$\_$12$\_$CN-1-2012-0016 and
 by the NSFC of China with Project Nos. 11322546, 11435004.

\end{document}